2009

# Mortality and Longevity Valuation

A Quantitative Approach

Where we present a brief discussion on methods for valuing longevity risk.

Louis Mello
ZHSquared
10/28/2009

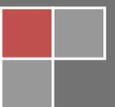

# Table of Contents

**Chaos is the origin of the universe. Mathematics is its language.**

## 0 - The Problem Stated

It is clear that the asset class comprised of life insurance related products that carry longevity risk cannot be valued based on typical financial measures. Specifically, the returns and yields so commonly bantered about amongst the 'players' have absolutely no meaning, because they are based on a flawed assumption that permeates the entire process: **the FALSE premise that one knows when an individual is going to die**. Nothing could be farther from the truth, nothing could be more difficult (nigh impossible) to predict, and this extremely dangerous assumption creates the illusion that the numbers spun by these 'players' can be used to make decisions. Recent events in the Mortgage Market (for those of shorter memory) as well as countless past collapses attest to just how dangerous these assumptions are.

It is clear, then, that the only number that really matters in this asset class is volatility-of-life-expectancy (V.O.L.E). Because the experience that is publicly available is restricted to Official Census Data, special studies such as TOAMS, and the mortality tables prepared by the SOA, there are some caveats that must be observed.

1) These data are based on large groups of lives and are not statistically suited for predicting when one individual is going to die.
2) These data are not necessarily demographically or geographically compatible with the probabilities that should be associated with an individual life.

Yet another fallacy is the LE provider number. Any commercial venture that claims to use medical information to estimate life expectancy, yet, whose estimates generally lie suspiciously close to those that one would get from the VBT 2008 gives cause for concern. Essentially, what one glimpses from these reports is that they know little more than we do about the health and life expectancy of the individuals they purport to have examined. So, to rely heavily on LE reports is to assume the same dangerous behavior mentioned above. These are, at best, merely another data point.

**So, the notion of variability in time of death is essential to any understanding of this asset.**

To this end we have devised a method to estimate this variance which is very straightforward and is founded on solid mathematical ground.

## 1- Simulation

Using the VBT 2008 we input age, gender and risk class as well as implied multiplier and improvement factor to simulate hundreds of thousands of times of death using a straightforward Monte-Carlo method. For any given individual we estimate the mode of life expectancy (the reason for the use of the mode is that the mean can vary widely from the mode – the most frequent occurrence – when the distribution is fat-tailed and skewed) and the maximum time of death, as well as the actuarial life expectancy.

This is given by[1]:

$$\overset{\circ}{e}_x = E[T(x)] = \int_0^\infty t \, {}_tp_x \, \mu(x+t) dt \qquad (1.1)$$

$$= \int_0^\infty t \, d_t(-{}_tp_x) \qquad (1.2)$$

$$= t(-{}_tp_x)\Big|_0^\infty + \int_0^\infty {}_tp_x \, dt \qquad (1.3)$$

The existence of a $E[T(x)]$ implies the limit $\lim_{t\to\infty} t(-{}_tp_x) = 0$, so we can write:

$$\overset{\circ}{e}_x = \int_0^\infty {}_tp_x \, dt. \qquad (1.4)$$

This is defined as the ***complete-expectation-of-life*** and constitutes the standard actuarial description.

With the values of $\overset{\circ}{e}_x$ and $\max\left[\overset{\circ}{e}_x\right]$ (the latter obtained via simulation of time of death) we can write:

$$\sigma = \frac{\max\left[\overset{\circ}{e}_x\right] - \overset{\circ}{e}_x}{\max\left[\overset{\circ}{e}_x\right]}$$

$$\sigma = 1 - \frac{\overset{\circ}{e}_x}{\max\left[\overset{\circ}{e}_x\right]} \qquad (1.5)$$

Where $\sigma$ is defined as the volatility - V.O.L.E.[2]

Now that we have defined volatility, the key ingredient in valuing any life insurance product, we can proceed to answer the next question:

How much is the mortality component of this contract worth? How much is it worth for me to have the right to a windfall on early death and a possible loss if the individual lives longer than expected?

---

[1] In *Bowers et al.* <u>Actuarial Mathematics</u> – SOA - 1997

[2] See Appendix II for the full mathematical rationale behind this assumption.

| | CF 1 | CF 2 | CF 3 | CF 4 | CF 5 | CF 6 | CF 7 | CF 8 |
|---|---|---|---|---|---|---|---|---|
| IRR | 295.1170% | 84.0403% | 42.7465% | 26.7200% | 18.5144% | 13.6855% | 9.6556% | 5.0508% |
| | $ (958,446.94) | $ (958,446.94) | $ (958,446.94) | $ (958,446.94) | $ (958,446.94) | $ (958,446.94) | $ (958,446.94) | $ (958,446.94) |
| | $ 3,786,986.96 | $ (739,049.07) | $ (739,049.07) | $ (739,049.07) | $ (739,049.07) | $ (739,049.07) | $ (739,049.07) | $ (739,049.07) |
| | | $ 4,606,486.64 | $ (802,335.09) | $ (802,335.09) | $ (802,335.09) | $ (802,335.09) | $ (802,335.09) | $ (802,335.09) |
| | | | $ 5,439,049.34 | $ (841,135.87) | $ (841,135.87) | $ (841,135.87) | $ (841,135.87) | $ (841,135.87) |
| | | | | $ 6,329,574.37 | $ (887,803.04) | $ (887,803.04) | $ (887,803.04) | $ (887,803.04) |
| | | | | | $ 7,268,100.10 | $ (939,592.16) | $ (939,592.16) | $ (939,592.16) |
| | | | | | | $ 8,264,355.17 | $ (1,002,515.01) | $ (1,002,515.01) |
| | | | | | | | $ 9,000,000.00 | $ (1,050,658.79) |
| | | | | | | | | $ 9,000,000.00 |

**Figure 1 - Demonstrates the returns on cash flows given different assumptions of time to death.**

This represents the only true value in these contracts.

In order to answer this question we revert to derivatives theory. In essence this translates to a call option (for which one pays a premium) to have the right to this very uncertain return.

Simply and clearly put[3]:

"The fair value of an option is the present value of the expected payoff at expiry

under a risk-neutral random walk for the underlying…"

A short word on risk neutrality: the random walk of the underlying must have a drift rate that is equal to the risk-free interest rate. This assumption means that you must be able to hedge the underlying at the risk-free rate. One can reasonably assume this as possible given that a corresponding annuity at the risk-free rate (at minimum) could be used to offset the intrinsic longevity risk.

## 2 - Asset Value[4]

Following Stone and Zissu[5] the valuation of a life settlement is given by the present value of the premiums, $-P$ and the present value of the Death Benefit, $B$. If one assumes a flat yield curve $r$ and a period ending in $t$ then one can write:

$$LSV = -P \left[ \frac{1}{(1+r)^1} + \frac{1}{(1+r)^2} \cdots \frac{1}{(1+r)^t} \right] + \frac{B}{(1+r)^t} \quad (1.6)$$

---

[3] Paul Wilmott – *Quantitative Finance* - 2007

[4] For a more complete summary see Appendix I.

[5] Journal of Alternative Investments, Volume Fall 2008, Number 25071

(1.6) can be rewritten as:

$$LSV = -P\left[\frac{1}{r} - \frac{a^t}{r}\right] + B \cdot a^t \quad (1.7)$$

where

$$a = \frac{1}{(1+r)}.$$

Rearranging (1.7) yields:

$$LSV = a^t\left[\frac{P}{r} + B\right] - \frac{P}{r} \quad (1.8)$$

The first derivative of (1.8) wrt $t$ is:

$$\frac{dLSV}{dt} = \left[\frac{P}{r} + B\right] a \cdot t \cdot \log(a) \quad (1.9)$$

Multiplying by $t$ and dividing by $LSV$ we can find *le duration*:

$$le\ duration = \left[\frac{(t \cdot a^t \cdot (P + r \cdot B) \cdot \log(a))}{(a^t \cdot (P + r \cdot B) - P)}\right] \quad (1.10)$$

When the *le duration* is negative (i.e. the *LSV* is positive) then we can say that the longer the insured lives beyond his LE = $t$, the more value the policy will lose.

This assumes level premia and death benefit. For the purpose of our study, the same logic applies only the values for $P$ and $B$ are replaced with vectors containing the values over time for these inputs.

Hence, we have found the asset value and the payoff for our option model.

# 3- Findings on the Random Walk

First and foremost, we have, through an exhaustive process of distribution fitting, arrived at the conclusion that the simulated times of death, using the VBT 2008 ANB for both genders and risk classes, is not stationary with respect to age. This means that as we simulate older people, the distribution becomes more fat-tailed and right-skewed. This obviously means higher volatility. More specifically, one has to refit the distribution for each case because there is no mathematical relationship, as of yet, for the parameters at one age as compared to those of another age. Noteworthy is the fact that we do not have (as do insurance companies and large portfolios) the luxury of the Law of Large Numbers.

This has resulted in the need to fit the simulated times of death to a Lévy Stable distribution which is a general class of distributions under which the Gaussian is a special case.

| Age | α MLE |
|-----|-----------|
| 60  | 2.0000000 |
| 61  | 1.9999000 |
| 62  | 2.0000000 |
| 63  | 1.9999000 |
| 64  | 2.0000000 |
| 65  | 1.9999000 |
| 66  | 2.0000000 |
| 67  | 1.9999000 |
| 68  | 2.0000000 |
| 69  | 1.9999000 |
| 70  | 2.0000000 |
| 71  | 1.9999000 |
| 72  | 2.0000000 |
| 73  | 2.0000000 |
| 74  | 2.0000000 |
| 75  | 2.0000000 |
| 76  | 1.9999000 |
| 77  | 1.9999000 |
| 78  | 2.0000000 |
| 79  | 1.9862000 |
| 80  | 1.9746000 |
| 81  | 2.0000000 |
| 82  | 1.9038000 |
| 83  | 1.8843000 |
| 84  | 1.8382000 |
| 85  | 1.7379000 |
| 86  | 1.6682000 |
| 87  | 1.6016000 |
| 88  | 1.5423000 |
| 89  | 1.5104000 |
| 90  | 1.5177000 |
| 91  | 1.4062000 |
| 92  | 1.2663000 |
| 93  | 1.2453000 |
| 94  | 1.1706000 |
| 95  | 1.1701000 |

**Figure 2 Illustrates the changes in the shape of the distribution given changes in age.**

This table demonstrates the change in the distribution observed with advancing age. The values of $\alpha$ were estimated using Nolan's[6] Maximum Likelihood Estimator.

---

[6] John P. Nolan – *Maximum Likelihood Estimation and Diagnostics for Stable Distributions* – Levy Processes Theory and Applications – Birkhäuser (2001)

The characteristic function for Lévy Stable distributions follows the standard formulation[7]:

$$\log(f(t)) = i\delta t - \gamma |t|^\alpha \cdot \left(\left(1 + i\beta \frac{t}{|t|}\right) \tan\left(\alpha \frac{\pi}{2}\right)\right) \qquad (1.11)$$

The four parameters are:

$\alpha =$ the kurtosis of the distribution as well as the fatness of the tails. It can range from 0 to $a \leq 2$.

$\beta =$ is the measure of skewness. When $\beta = +1$ the distribution is right-skewed which is usually the case in the distributions we have looked at. The reason is quite simple: there is always a 1 period minimum (i.e. the individual can only die in the future, and the shortest period in the future is 1). All of the risk lies to the right, that is, in the extension of life expectancy.

$\gamma =$ is the scale parameter, that is, the adjustment made for different time horizons in the underlying data, e.g. a change from yearly to monthly.

$\delta =$ is the location parameter.

Notice that if we take (1.11) and plug in $\alpha = 2$, $\beta = 0$, $\gamma = 1$, $\delta = 1$ we get:

$$\log(f(t)) = i\mu t - \frac{\sigma^2}{2} t^2 \qquad (1.12)$$

where $\mu =$ mean and $\sigma^2 =$ variance.

This is the characteristic function of the Normal (Gaussian) distribution and it proves what we said above about the Gaussian being a special case of the Lévy Stable class of distributions.

One important characteristic of the non normal class of distributions is with respect to sample variance. In the case of $1 \leq \alpha < 2$ variance becomes undefined, or infinite. Variance is only finite and stable when $\alpha = 2$ which is not necessarily the norm. It is for this reason that we have described above an alternate method for estimating volatility.

## 4 - Log Normal and Lévy Stable Distributions

McCulloch[8] and Zolotarev[9] worked independently on a sometimes forgotten aspect of Lévy Stable Distributions. When the skewness parameter, $\beta$, is at either extreme (+1 or -1), the affected tail loses

---

[7] Paul Lévy – *Theorie de l'Addition des Variables Aleatoires* – Paris, Gauthier-Villars (1937)

[8] J.H. McCulloch *The Value of European Options with Log Stable Uncertainty* – Working Paper, 1985

its Paretian traits and has a steeper declining gradient than even the normal distribution. The opposite tail becomes longer and fatter so the distribution resembles a log normal – unimodal with a long positive tail and a short negative tail.

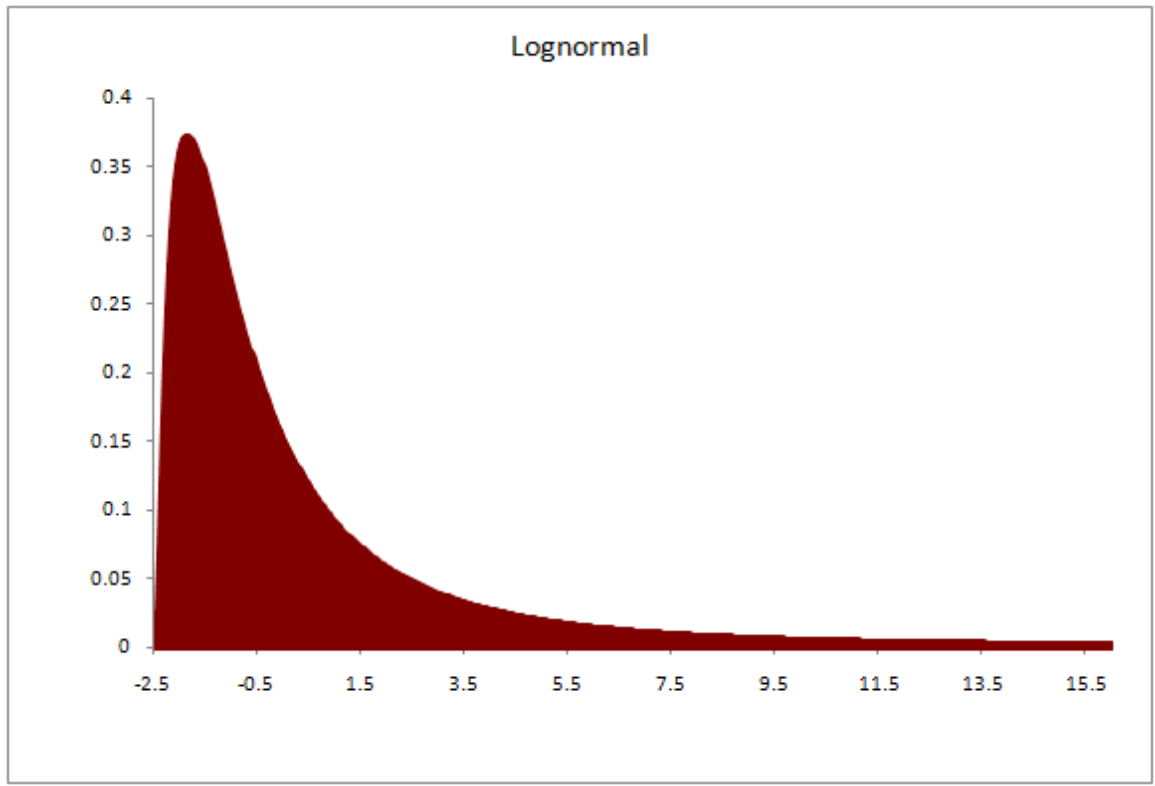

**Figure 3  Typical Log Normal Distribution**

As a result, we can model the stochastic path with a log normal stochastic differential equation that has an exact and simple time stepping algorithm.

$$d(\log S) = \left(r - \frac{1}{2}\sigma^2\right)dt + \sigma \, dX \qquad (1.13)$$

After integration, this yields:

$$S(t) = S(0)\exp\left(\left(r - \frac{1}{2}\sigma^2\right)t + \sigma\int_0^t dX\right) \qquad (1.14)$$

---

[9] V.M.Zolotarev *On Representation of Stable Laws by Integrals* – Selected Transactions in Mathematical Statistics and Probability, Vol. 6. Providence R.I.: American Mathematical Society, 1966. (Russian original, 1964)

And transforming into a time step, $\delta t$

$$S(t+\delta t) = S(t) + \delta S = S(t)\exp\left[\left(r - \frac{1}{2}\sigma^2\right)\delta t + \sigma\sqrt{\delta t}\varepsilon\right] \quad (1.15)$$

The beauty of this solution is that one can simulate both path dependent and path independent processes because (1.14) is exact, and hence the time steps do not have to be small. In fact, one could model a path independent process, such as a European option, in one large jump of time step $T$.

One can also randomize time of death by making $t$ of $\delta t$ a random integer taken from a skewed normal distribution.

In order generate $\varepsilon$, normally distributed random numbers, we employed the Box-Muller method which is defined as:

$$\begin{aligned} y_1 &= \sqrt{-2\log(R_1)} \cdot \cos(2\pi R_2) \\ y_2 &= \sqrt{-2\log(R_1)} \cdot \sin(2\pi R_2) \end{aligned} \quad (1.16)$$

Where $R_1$ and $R_2$ are uniform deviates. $y_1$ and $y_2$ are Gaussian deviates.

## 5 - Monte-Carlo Simulation

Monte-Carlo simulations operate on very easily understood principles. The math employed is not complicated. Correlations can be easily modeled. It is a simple matter of increasing the number of simulations in order to improve the accuracy of the results. The models are easily adapted, and the results remain robust (one eliminates the oft encountered error of assuming away the true risk inherent in the modeling process). Complex path dependency is easily incorporated. The methodology is widely documented and accepted not only amongst financial practitioners but in the scientific community as well.

# Appendix I - Duration Calculations and Policy Present Value

What follows is the condensed mathematics behind the Duration calculation as proposed in ***Securitized Senior Life Settlements' Macaulay Duration and Longevity Risk*** By Ortiz, Stone and Zissu.

The present value of a life settlement policy is given by:

$$P = -p \sum_{i=1}^{n} \left[ \frac{1}{(1+r)^i} \right] + \frac{DB}{(1+r)^n} \quad (1.17)$$

Where:

$-p$ = the premium paid per period

$n$ = the number of periods

$DB$ = the death benefit

$r$ = the discount rate

The general formula for Macaulay Duration, $D$, can be written as:

$$D = \sum_{i=1}^{t} \frac{i \cdot cf_i}{(1+r)^i} / P \quad (1.18)$$

Where:

$i$ = the period at which respective cash flow ($cf$) is paid

$r$ = the discount rate

$P$ = the present value at the time $D$ is calculated

$cf_i$ = the cash flow at time $i$

Since the cash flows to and from a life settlement contract are the periodic premia (outflows) and the death benefit (inflow) at time $t$ we can represent Macaulay Duration for a life settlement contract as:

$$D = \left[\sum_{i=1}^{t} \frac{p_i}{(1+r)^i} - \frac{t \cdot DB}{(1+r)^t}\right] / P \tag{1.19}$$

In order to simplify the computation of the first derivative of (1.19) we assign $a = \dfrac{1}{1+r}$.

Now we isolate the term $f(t) = \sum_{i=1}^{t} p_i (a)^i$.

Simple algebra tells us that  (1.20)  $\sum_{i=k}^{n} b^i = \dfrac{(b^{n+1} - b^k)}{b-1}$  where $b \in \mathbb{R}$ and $k < n$

If we rewrite $f(t)$ as $f(t) = p \sum_{i=1}^{t} i(a)^i$ then (1.20) can be applied to every segment of the expansion of $f(t)$ and after some algebraic manipulation we see that

$$f(t) = \left[p / (a-1)^2\right] \cdot \left[t \cdot a^{t+2} - (t+1) \cdot a^{t+1} + a\right] \tag{1.21}$$

Where, as before, $a = \dfrac{1}{1+r}$

And $p$ and $r$ are both constant with respect to $t$.

Now, we can add (1.21) to (1.19) which yields

$$D = \left(\frac{1}{P}\right)\left\{t \cdot a^t \left[a \cdot p / (a-1) - DB\right] - a^t \left[a \cdot p / (a-1)^2\right] + a \cdot p / (a-1)^2\right\} \tag{1.22}$$

In order to obtain the derivative of (1.22) wrt time we change the form to

$$D = \left(\frac{1}{P}\right) \cdot \left\{t \cdot a^t \cdot \left[C \cdot (a-1) - DB\right] - a^t \cdot C + C\right\} \tag{1.23}$$

Where $a, DB, P$ and $C$ are constants and $C = \dfrac{a \cdot p}{(a-1)^2}$.

Now we take the derivative of (1.23) wrt time whose solution is

$$D' = \left(\frac{a^t}{P}\right) \cdot \left\{\left[t\left[C(a-1) - DB\right]\log(a)\right] + \left[C \cdot (a-1) - DB - C\left[\log(a)\right]\right]\right\} \tag{1.24}$$

Using equation (1.24) we can now obtain the solution for $D=0$ which translates to the point at which duration does not change given a shift (upward or downward) in life expectancy.

If we note that $\dfrac{a^t}{P}$ can never equal zero given that $a = \dfrac{1}{1+r}, \; >0$ then the duration wrt $t$ equals zero when:

$$t(C \cdot (a-1) - DB) + C \cdot (a-1) - DB - C \cdot (\log(a)) = 0 \qquad (1.25)$$

Replacing $C$ and solving for $t$ yields:

$$t^* = \left[-1/\log(a)\right] + a \cdot p / \left[a \cdot p \cdot (a-1) - DB \cdot (a-1)^2\right]$$

# Appendix II[10] - Derivation of Mortality Model

Given a population of one individual, we know that at some point in the future he will die.

Let the random variable $X(t)$ denote the number of live persons in this population. Then $X$ has a sample space $\{0,1\}$; i.e. the two possible states of the population are $i=0$ and $i=1$, and the matrix below describes this:

$$\mathbb{R}(t) = \begin{pmatrix} r_{00}(t) & r_{01}(t) \\ r_{10}(t) & r_{11}(t) \end{pmatrix}$$

Let us then derive an explicit expression for the entries of this matrix.

If our sole individual were dead at time $t=0$ then he would remain so.

Hence, $\Pr(X(t)=0 \mid X(0)=0)=1$, $\Pr(X(t)=1 \mid X(0)=0)=0$. Thus, $r_{00}(t)=1$ and $r_{01}(t)=0$. However, the person is alive at time $t=0$, such that $r_{00}$ and $r_{01}$ have no bearing on his future. Then, because the last row of the matrix $\mathbb{R}$ must sum to 1, $N=1$, we have $r_{10}(t)+r_{11}(t)=1$. Hence,

$$\mathbb{R}(t) = \begin{pmatrix} 1 & 0 \\ 1-r_{11}(t) & r_{11}(t) \end{pmatrix} \tag{1.26}$$

Since the individual is alive initially we have $\pi(0)=(0,1)^T$. This allows us to write the system distribution vector $\pi(t)$ as:

$$\pi(t)^T = (0,1)\mathbb{R}(t) = \left(1-r_{11}(t), r_{11}(t)\right), \tag{1.27}$$

and of course $\pi$ is completely determined as long as we know $r_{11}(t)$.

In order to find $r_{11}(t)$ we shall employ a differential equation that requires an initial condition. This is clearly given by the fact that if the person is alive at time $t=0$ then he cannot also be dead at time $t=0$, however, he could be dead at time $t=\varepsilon$ for any value of $\varepsilon>0$, however small. Hence,

$$r_{11}(0) = \Pr(X(0)=1 \mid X(0)=1) = 1 \tag{1.28}$$

---

[10] Mesterton-Gibbons, Michael – *A Concrete Approach to Mathematical Modelling* – J.

Let $U_1$ be the event that the person is alive at time $t$, i.e. that $X(t)=1$. Let $U_0$ be the event that he is dead at time $t$, i.e. that $X(t)=0$. Let $U$ be the event that that he is alive at the later time $t+\delta t$, i.e. that $X(t+\delta t)=1$ where $\delta t$ is infinitesimally small. Thus, $U_0$ and $U_1$ are associated with the random variable $X(t)$, and $U$ with the random variable $X(t+\delta t)$; but both random variables have the same sample space $\{0,1\}$ and belong to the same stochastic process. The person must be either alive or dead at time $t$ and so $U_0$ and $U_1$ scan all possible scenarios. If $N=1$ we have

$$\Pr[U] = \Pr[U|U_0] \cdot \Pr[U_0] + \Pr[U|U_1] \cdot \Pr[U_1] \tag{1.29}$$

But we know that $\Pr[U|U_0]=0$ because the person cannot be alive at $t+\delta t$ if he is dead at time $t$. Also, from (1.27) we have $\Pr[U] = \Pr(X(t+\delta t)=1) = r_{11}(t+\delta t)$; and $\Pr[U_1]\Pr(X(t)=1) = r_{11}(t)$. Moreover, because $\Pr[U|U_1]$ is the probability that the individual survives from $t$ to $\delta t$, $1-\Pr[U|U_1]$ is obviously the probability that he dies in the interval $[t, t+\delta t]$, which can be made equivalent to $d_1 \delta t + o(\delta t)$. So, (1.29) reduces to

$$r_{11}(t+\delta t) = (1 - d_1 \delta t + o(\delta t)) r_{11}(t). \tag{1.30}$$

Rearranging and dividing by $\delta t$, we get

$$\frac{r_{11}(t+\delta t) - r_{11}(t)}{\delta t} = -d_1 r_{11}(t) + \frac{o(\delta t)}{\delta t}. \tag{1.31}$$

Taking the limit as $\delta t \to 0$ we have

$$\frac{d}{dt}\{r_{11}(t)\} = -d_1 r_{11}(t). \tag{1.32}$$

Solving the ODE and using (1.27) and (1.28) we will arrive at

$$\pi_1(t) = r_{11}(t) = r_{11}(0)e^{-d_1 t} = e^{-d_1 t} \tag{1.33}$$

$$\pi_0(t) = r_{10}(t) = 1 - r_{11}(t) = 1 - e^{-d_1 t}. \tag{1.34}$$

Thus, we verify that the probability that the state of the population is 1 – i.e. the probability that the person does not die – decays to zero exponentially as $t \uparrow \infty$. This is simply the probabilistic analogue of a natural decay process.

Now, in order to satisfy the Markov property we let $G$ denote the time until death in the individual's life. The cumulative distribution function of $G$ is given by

$$\Pr(G > t) = e^{-d_1 t} \tag{1.35}$$

Or

$$\Pr(G \leq t) = 1 - e^{-d_1 t}. \tag{1.36}$$

We should readily recognize this as the *exponential distribution*, the only one with the property such that

$$\Pr(G > t + s \mid G > s) = \Pr(G > t) \tag{1.37}$$

for all $t, s \geq 0$. It is clear that (1.37) is a statement of Markov's no-memory property.

The probability density function of the exponential distribution can be written

$$f(t) = \frac{d}{dt}\{\Pr(G \leq t)\} = d_1 e^{-d_1 t}, \quad 0 \leq t \leq \infty \tag{1.38}$$

and as a result, the mean, $\mu$, and the variance, $\sigma^2$, of the distribution are given by

$$\mu = \int_0^\infty t f(t) dt = \frac{1}{d_1} \tag{1.39}$$

$$\sigma^2 = \int_0^\infty (t - \mu)^2 f(t) dt = \frac{1}{d_1^2}. \tag{1.40}$$

**Figure 1 A Typical Cumulative Exponential Distribution**

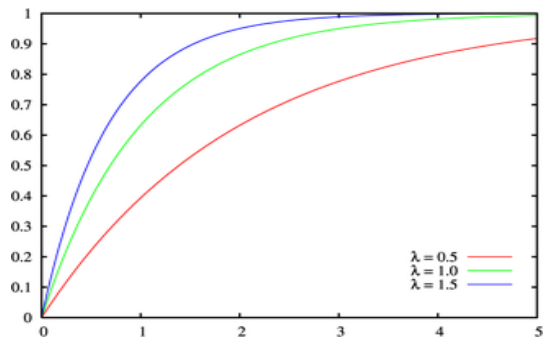

From (1.39) we can find the death rate:

$$d_1 = \frac{1}{\text{Mean Time Until Transition}} \tag{1.41}$$

## Appendix III - Variance from the Life Table

A continuous failure time, $T$, must be associated with a discrete failure time $\tilde{T}$. A discrete failure will occur at an integer time, $k$, given that failure under $T$ has occurred in the interval $(k-1, k]$. This means:

$$\tilde{T} = [T]+1, \quad \text{where } [\cdot] \text{ denotes the greatest integer function.}$$

Thus, let $f$ and $s$ denote the density function and survival function, respectively, of $T$. Let $\tilde{f}$ and $\tilde{s}$ represent the corresponding functions for $\tilde{T}$. We can then see that for all positive integers $k$,

(2.1)
$$\tilde{f}(k) = \int_0^1 f(k-1+t)\,dt = s(k-1) - s(k)$$
$$\tilde{s}(k) = s(k).$$

Because we may not be able to observe failure times continuously, i.e., we know that failure has occurred at time $k$ but can only be sure that it actually took place between $k-1$ and $k$. Thus, we observe $\tilde{T}$ but would like to be able to speak to $T$. It is clear that one cannot do this directly, and as a result, some form of approximation is needed. A simple method, consistent with (2.1) is simply to assume that, for all non-negative integers $k$, and any time $t$ in the interval $(0,1)$,

(2.2)
$$f(k+t) = \tilde{f}(k+1).$$

This is commonly called the *standard approximation*. In fact, it underlies the UDD assumption for the random variable $T(x)$, which is clear from the equivalent formulation in terms of the survival function:

(2.3)
$$s(k+t) = (1-t)\cdot s(k) + t\cdot s(k+1),$$

for any integer $k$ and $0 < t < 1$. As proof of this equivalence, given (2.3) one can obtain (2.2) by differentiation w.r.t. $t$. Conversely, given (2.2), one can show that:

(2.4)
$$s(k+t) = s(k) - \int_0^t f(k+r)\,dr$$
$$= s(k) - \int_0^t \tilde{f}(k+1)\,dr = s(k) - t\cdot \tilde{f}(k+1)$$
$$= s(k) - t\cdot[s(k) - s(k-1)].$$

In order to estimate moments we must introduce the random variable

$$R = \tilde{T} - T,$$

which represents the duration from the time of failure until the end of the year failure. Given a value $r$ in the interval $(0,1)$, consider the probability that $R \leq r$ given that $\tilde{T} = k$. What we ask is the probability that $T$ takes on a value between $k - r$ and $k$. Using the standard approximation, we can write:

$$\text{(2.5)} \quad \frac{1}{\tilde{f}(k)} \int_{k-r}^{k} f(t)dt = \frac{1}{\tilde{f}(k)} \int_{k-r}^{k} \tilde{f}(t)dt = r.$$

Hence, from (2.5) we see that $R$ is independent of $\tilde{T}$ and has a uniform distribution on the interval $(0,1)$. Using the standard calculations w.r.t. the uniform distribution, we can calculate

$$\text{(2.6)} \quad E(T) = E(\tilde{T}) - E(R) = E(\tilde{T}) - \frac{1}{2},$$

$$\text{(2.7)} \quad \text{Var}(T) = \text{Var}(\tilde{T}) + \text{Var}(R) = \text{Var}(\tilde{T}) + \frac{1}{12},$$

where $E$ denotes expectation and $\text{Var}$ denotes variance.

Now, once we have found life expectancy (2.6), we can proceed to the estimation of the second moment. Let $t$ be the time as measured on the life table where $\omega$ is the ending year. Variance can be written as:

$$\text{(2.8)} \quad \sigma^2 = \sum_{i=1}^{\omega} \left[ \left( \overset{\circ}{e} - i \right)^2 \cdot \Pr(d_i) \right] + \frac{1}{12}$$

where $\overset{\circ}{e}_x = E[T(x)] =$ complete life expectancy (1.4) and $\Pr(d_i) =$ probability of death at time $i$.

# APPENDIX IV - Pricing Life Settlement as an American Option by way of Finite Differencing

By definition an American Option can be exercised at any time prior to expiration. Thus, valuing an American Option entails the solution of a dynamic stochastic optimization problem. This means that at each instant in time one must decide if it is optimal or not to exercise the option.

Formally, the price of an American Option can be written:

$$\max_{\tau} \hat{E}\left[e^{-r\tau}\Phi(S_\tau)\right] \quad (2.9)$$

where $\Phi$ is the payoff and expectation is taken under a risk-neutral measure. Obviously, $\tau$ is the stopping time. The stopping time, in this context, may be understood as the time at which one exercises. This, then, is a random variable that depends solely on the history of price thus far.

It is clear that early exercise will not occur if the option is out-of-the-money. For a put, one would not exercise the option if $S(t) > K$. However, even if $S(t) < K$ it may still be more advantageous to wait for a "better" time to exercise. In other words, the option should be "enough" in-the-money. "Enough" here will generally depend on how close one is to expiration. Unmistakably, the closer to expiration, the more prone one would be to early exercise.

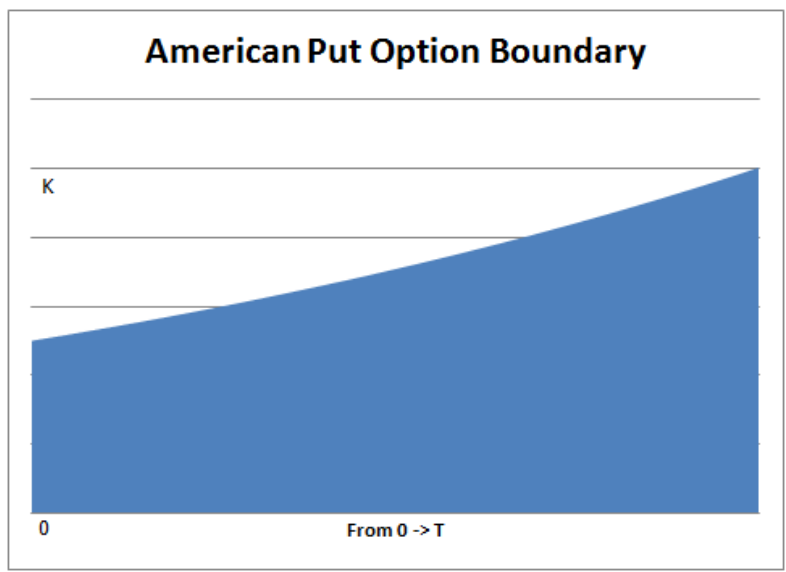

**Figure 4**

In Figure 1 one can visualize how anywhere inside the shaded area is a feasible exercise point.

Finding the optimal boundary is what makes pricing American Options so much more difficult than pricing European Options: one cannot simply compute an expected value. The solution to this problem, within the framework of partial differential equations, becomes one of *free boundary*. Another possible solution lies in finite differencing, which, in essence, boils down to comparing the intrinsic and continuation values in order to make a decision.

## Introduction to the Finite Difference Method

Finite difference approaches to solving PDEs are based on the simple idea of approximating each partial derivative by a difference quotient. This will transform the functional equation into a set of algebraic equations. As is the case for most numerical algorithms, the starting point is a finite series approximation. Under suitable continuity and differentiability hypotheses, Taylor's theorem states that a function $f(x)$ may be represented as

$$f(x+h) = f(x) + hf'(x) + \frac{1}{2}h^2 f''(x) + \frac{1}{6}h^3 f'''(x) + \cdots \qquad (2.10)$$

Neglecting the terms of order $h^2$ and higher, we get

$$f'(x) = \frac{f(x+h) - f(x)}{h} + O(h). \qquad (2.11)$$

It is clear that (2.11) represents a *forward* approximation for the derivative; indeed, the derivative is simply defined as a limit of the difference quotient above as $h \to 0$.

Similarly, one could write

$$(2.12) \qquad f(x-h) = f(x) - hf'(x) + \frac{1}{2}h^2 f''(x) - \frac{1}{6}h^3 f'''(x) + \cdots$$

which would yield

$$(2.13) \qquad f'(x) = \frac{f(x) - f(x-h)}{h} + O(h).$$

Whence one would obtain a *backward* approximation.

In either case one finds a truncation error of order $O(h)$. A better approximation is found by subtracting (2.12) from (2.10) and after rearranging

$$(2.14) \qquad f'(x) = \frac{f(x+h) - f(x-h)}{2h} + O(h^2).$$

This translates to the *central* or *symmetric* approximation, and for a small $h$ it is a better approximation, since the truncation error is $O(h^2)$.

This rationale may also be applied to higher order derivatives. To cope with the Black and Scholes equation, one must approximate second-order derivatives as well. This may be attained by adding equations (2.10) and (2.12) which yields

(2.15) $$f(x+h)+f(x-h)=2f(x)+h^2 f''(x)+O(h^4)$$

and rearranging

(2.16) $$f''(x)=\frac{f(x+h)-2f(x)+f(x-h)}{h^2}+O(h^2).$$

In order to apply this concept to a PDE involving a function $\phi(x, y)$, it is natural to set up a discrete grid of points of the form $(i\,\delta x, j\,\delta y)$, where $\delta x$ and $\delta y$ are steps for discretization, and then seek out values of $\phi$ on this grid. It is customary to employ standard grid notation:

(2.17) $$\phi_{ij}=\phi(i\,\delta x, j\,\delta y)$$

This process can provide a set of algebraic equations with a relatively simple solution.

A possible difficulty may arise from the boundary conditions. If the equation is defined over a rectangular domain in the $(x, y)$ space the grid is easily set up such that the boundaries are on the grid. However, there are cases where this will not be so simple. Nonetheless, one expects that for $\delta x, \delta y \to 0$ the solution of this set converges to the solution of the PDE.

## Basic Black & Scholes

We must first introduce a basic set of equations that model the behavior of a class of derivatives. In particular, we model derivatives that are described by so-called initial boundary value problems of parabolic type. Thus, consider the general parabolic equation

(3.1) $$Lu \equiv \sum_{i,j=1}^{n} a_{ij}(x,t)\frac{\partial^2 u}{\partial x_i \partial x_j}+\sum_{j=1}^{n} b_j(x,t)\frac{\partial u}{\partial x_j}+c(x,t)u-\frac{\partial u}{\partial y}=f(x,t)$$

where the functions $a_{ij}, b_j, c$ and $f$ are real and take finite values, $a_{ij}=a_{ji}$ and

(3.2) $$\sum_{i,j=1}^{n} a_{ij}(x,t)a_i a_j > 0 \text{ if } \sum_{j=1}^{n} a_j^2 > 0$$

In (3.2) the variable $x$ is a point in n-dimensional space and $t$ is considered to be a positive time value. Equation (3.1) is the general equation that describes the behavior of many types of derivatives. In the one dimensional case $(n=1)$ it reduces to the Black & Scholes equation (see Black 1973)

$$(3.3) \qquad \frac{\partial V}{\partial t^*} + \sum_{j=1}^{n}(r-D_j)\cdot S_j \frac{\partial V}{\partial S_j} + \frac{1}{2}\sum_{i,j=1}^{n}\rho_{ij}\cdot\sigma_i\cdot\sigma_j\cdot S_i\cdot S_j\cdot\frac{\partial^2 V}{\partial S_i \partial S_j} = r\cdot V$$

where $V$ is the derivative type, $S$ is the underlying asset, $\sigma$ is the constant volatility, $r$ is the interest rate and $D$ is a dividend.

Equation (3.1) will typically have an infinite number of solutions. Hence, to reduce that number to one, we must introduce constraints. So, we define the initial condition and boundary conditions for (3.1). We must, then, define the space in which equation (3.3) is assumed valid. Since the equation has a second order derivative in $x$ and a first order derivative in $t$ we expect that a unique solution can be found by defining two boundary conditions and one initial condition. In general there are three types of boundary conditions associated with (3.1):

1. First boundary value problem (Dirichlet Problem);
2. Second boundary value problem (Neumann, Robins problems);
3. Cauchy problem.

The first boundary value problem deals with the solution of (3.1) in a domain $D = \Omega X(0,T)$ where $\Omega$ is a bounded subset of $\mathbb{R}^n$ and $T$ is a positive number. In this case we seek to find a solution to (3.1) satisfying the conditions

$$(3.4) \qquad \begin{aligned} u\,|_{t=0} &= \varphi(x) \quad \text{initial condition} \\ u|_{\Gamma} &= \psi(x,t) \quad \text{boundary condition} \end{aligned}$$

where $\Gamma$ is a subset of $\Omega$. The boundary conditions in (3.4) are sometimes called Dirichlet boundary conditions. The second boundary value problem is similar to (3.4) except that, instead of giving the value of $u$ on the boundary $\Gamma$, the directional derivatives are included, as can be verified by

$$(3.5) \qquad \left(\frac{\partial u}{\partial \eta} + a(x,t)u\right)|\Gamma = \psi(x,t)$$

In this case, $a(x,t)$ and $\psi(x,t)$ are known functions of $x$ and $t$, and $\frac{\partial}{\partial \eta}$ denotes the derivative of $u$ with respect to the outward normal $\eta$ at $\Gamma$. A special case of (3.5) is when $a(x,t)\equiv 0$; then (3.5) represents the Neumann boundary conditions. These usually occur when modeling certain put options. It is clear that (3.5) should be augmented by an initial condition similar to that in equation (3.4). Finally, the solution to the Cauchy problem for (3.1) in the strip $\mathbb{R}^n X(0,T)$ is given by the initial condition

(3.6)
$$u|_{t=0} = \varphi(x)$$

where $\varphi(x)$ is a given continuous function and $u(x,t)$ is a function that satisfies (3.1) in $\mathbb{R}^n X(0,T)$ and satisfies the initial condition (3.6). This problem allows negative values of the independent variable $x = (x_1, \ldots, x_n)$. A special case of the Cauchy problem can be seen in the modeling of one-factor European and American options (see Wilmott 1993) where $x$ is represented by the underlying asset $S$.

Boundary conditions are given by values $S = 0$ and $S = \infty$. For European options these conditions are:

(3.7)
$$C(0,t) = 0$$
$$C(S,t) \to S \quad \text{as} \quad S \to \infty$$

Here $C$ (the role played by $u$ in equation (3.1)) is the variable representing the price of the call option. For European put options, on the other hand, the boundary conditions are

(3.8)
$$P(0,t) = K \cdot e^{-r(T-t^*)}$$
$$P(S,t) \to 0 \quad \text{as} \quad S \to \infty$$

Here $P$ (the role played by $u$ in equation (3.1) is the variable that represents the price of the put option, $K$ is the strike price, $r$ is the risk-free rate, $T$ is the time to expiry and $t$ is the current time. In practice, we see many numerical solutions for European options by assuming a finite domain, that is one in which the right-hand boundary conditions in (3.7) and (3.8) are defined at large but finite values of $S$. For the remainder of this paper we shall designate $t = T - t^*$ as the independent time variable. Thus, be forewarned that we are solving initial boundary value problems rather than terminal boundary value problems. Hence, from this point on we assume the following canonical form for the $L$ operator in equation (3.1):

(3.9)
$$Lu \equiv \frac{\partial u}{\partial t} + \sigma(x,t) \cdot \frac{\partial^2 u}{\partial x^2} + \mu(x,t) \frac{\partial u}{\partial x} + b(x,t) u = f(x,t)$$

where $\sigma, \mu$ and $b$ are known functions of $x$ and $t$. For the moment, there are no boundedness or continuity conditions imposed on them.

## The Maximum Principle for Parabolic Equations

Let us now see how possible solutions of (3.1) and its one-factor equivalent, (3.9), behave in the desired domain. We shall look toward the continuous maximum principle that lays out the conditions under which the solution of (3.1) remains positive when its corresponding initial and boundary conditions are positive.

Let $D = \Omega X(0,T)$ and let $\bar{D}$ be its closure. The following results have been proven since 1962.

**Theorem 1.** *Assume that the function $u(x,t)$ is continuous in $D$ and assume that the coefficients in (3.9) are continuous. Suppose that $Lu \leq 0$ in $\bar{D} \setminus \Gamma$ where $b(x,t) < M$ ($M$ is a constant) and suppose, furthermore, that $u(x,t) \geq 0$ on $\Gamma$. Then*

$$u(x,t) \geq 0 \quad \text{in} \quad \bar{D}.$$

This theorem, simply put, states that positive initial and boundary conditions lead to a positive solution in the interior of the domain $D$. This will have far reaching consequences, as will be demonstrated.

**Theorem 2.** *Suppose that $u(x,t)$ is continuous and satisfies (3.9) in $\bar{D} \setminus \Gamma$ where $f(x,t)$ is a bounded function $(|f| \leq N)$ and $b(x,t) \leq 0$. If $|u(x,t)|_\Gamma \leq m$ then*

(3.10) $$|u(x,t)| \leq N \cdot t + m \quad \text{in} \quad \bar{D}.$$

We can sharpen the result of **Theorem 2** in the case where $b(x,t) \leq b_0 < 0$. In this case estimate (3.10) is replaced by

(3.11) $$|u(x,t)| \leq \max\left\{\frac{-N}{b_0}, m\right\}$$

*Proof*: Define the "barrier" function $w^\pm(x,t) = N_1 \pm u(x,t)$ where $N_1 = \max\left\{\frac{-N}{b_0}, m\right\}$. Then $w^\pm \geq 0$ and $Lw^\pm \leq 0$. By **Theorem 1** we can deduce that $w^\pm \geq 0$ in $\bar{D}$. The desired result follows.

The inequality (3.11) states that the growth of $u$ is bounded by its initial and boundary values. It is interesting to note that in the special case $b \equiv 0$ and $f \equiv 0$ we can deduce the following maximum and minimum principles for the heat equation and its variants.

**Corollary 1.** *Assume that the conditions of **Theorem 2** are satisfied and that $b \equiv 0$ and $f \equiv 0$. Then, the solution $u(x,t)$ takes its least and greatest values on $\Gamma$, that is*

$$m_1 = \min u(x,t) \leq u(x,t) \leq \max u(x,t) \equiv m_2$$

Now we may proceed to proving the positivity of solutions to (3.9) in unbounded domains, e.g. the case of American and European options.

**Theorem 3.** *(Maximum principle for the Cauchy problem)*

*Let $u(x,t)$ be continuous and bounded below in $H = \mathbb{R}^n X (0,T)$, that is $u(x,t) \geq -m, m > 0$. Furthermore, assume that $u(x,t)$ has continuous derivatives in $H$ up to second order in $x$ and first order in $t$ and that $Lu \leq 0$. Let $\sigma, \mu$ and $b$ satisfy*

$$|\sigma(x,t)| \leq M(x^2 + 1)$$
$$|\mu(x,t)| \leq M\sqrt{x^2 + 1}$$
$$b(x,t) \leq M$$

*Then, $u(x,t) \geq 0$ everywhere in $H$ if $u \geq 0$ for $T = 0$.*

We can apply **Theorem 3** to the one-factor Black & Scholes equation (3.3) in order to convince ourselves that the price of an option can never take negative values.

## Special Cases

Our specific area of interest is the one-factor generalized Black & Scholes equation with initial condition and Dirichlet boundary conditions.

Define $\Omega = (A, B)$ where $A$ and $B$ are two real numbers.

The formal statement then, is:

Find a function $u : D \to R^1 (D = \Omega X (0,T))$ such that:

(3.12) $$Lu \equiv \frac{\partial u}{\partial t} + \sigma(x,t) \cdot \frac{\partial^2 u}{\partial x^2} + \mu(x,t) \cdot \frac{\partial u}{\partial x} + b(x,t) \cdot u = f(x,t) \text{ in } D$$

(3.13) $$u(x,0) = \varphi(x), x \in \Omega$$

(3.14) $$u(A,t) = g_0(t), u(B,t) = g_1(t), t \in (0,T)$$

The initial-boundary value problem (3.12) - (3.14) is very general and it subsumes many specific cases from the risk literature (in particular, it is a generalization of the original Black & Scholes equation).

Usually the coefficients $\sigma(x,t)$ and $\mu(x,t)$ represent volatility (diffusivity) and drift (convection), respectively. Equation (3.12) is called convection-diffusion and it has been the object of much study. It serves as a model for many kinds of physical and economic phenomena. Research has shown that standard centered-difference schemes fail to approximate (3.12) - (3.14) in certain instances (see Farrell 2000).

Our goal is to show that the classical finite difference approach does not always produce good results and, consequently, devise a family of robust and accurate finite difference schemes for this problem.

Let us, then, investigate some special limiting cases in the system (3.12) - (3.14). First, when the function $\sigma(x,t)$ tends to zero as a function of $x$ or $t$ (or both). The classic Black & Scholes formulation assumes that volatility is constant, but in practice, this may not be true. For example, volatility may be time-dependent (see Wilmott 1993). In general, volatility may be a function of both time and the underlying variable. If the case were the former, then there exists an explicit solution. However, let us look at the exponentially declining volatility functions (see Deventer 1997) as given by

(3.15) $$\sigma(t) = \sigma_0 e^{-\alpha \cdot (T-t)}$$

where $\sigma_0$ and $\alpha$ are given constants.

Having described situations in which the coefficient of the second derivative in equation (3.12) can be small or tend to zero we now discuss what the mathematical implications are. This is quite important given that finite difference schemes must be robust enough to model the exact solution in all extreme cases. Setting $\sigma$ to zero in (3.12) leads to a formal first-order hyperbolic equation

(3.16) $$L_1 u \equiv -\frac{\partial u}{\partial t} + \mu(x,t) \cdot \frac{\partial u}{\partial x} + b(x,t) \cdot u = f(x,t)$$

Since the second derivative in $x$ is not present in (3.16) we conclude that only one boundary condition and one initial condition are needed to specify a unique solution (see Friedrichs 1958, Duffy 1977). However, we must answer the question: which boundary condition in (3.14) should be chosen? In order to answer this question we must define the characteristic lines associated with (3.16) (see Godounov 1979, Godounov 1973). These are defined as lines that satisfy the ODE

(3.17) $$\frac{dx}{dt} = -\mu$$

These lines have positive or negative slope depending on whether $\mu$ has negative or positive value. In general, it can be shown (see Friedrichs 1958) that to discover the "correct" boundary condition in (3.16) we must solve

(3.18) $$\begin{aligned} u(A,t) = g_0(t) \quad \text{if} \quad \mu < 0 \\ u(B,t) = g_1(t) \quad \text{if} \quad \mu > 0 \end{aligned}$$

We see that one of the boundary conditions in (3.14) is superfluous which leads to the renowned boundary layer phenomenon so common in fluid dynamics (see Vishik 1957 for a fundamental analysis of ODEs and PDEs in which the coefficient of the highest order derivative is small or tends to zero as $t \to \infty$).

So, we can see that as $\sigma \to 0$ we cannot satisfy both boundary conditions in (3.14). The presence of the resulting boundary layer causes major problems for classical finite difference methods because they are unable to cope with the approximation of the exact solution in the boundary layer itself. As usual, we say that system (3.12) - (3.14) is of singular perturbation type when either $\sigma$ is small or $\mu$ is large. We can also, sometimes, say that the system is convection-dominated in the latter case.

Another special case in (3.12) is when $\mu \to 0$ or is zero. The resulting equation is similar to the heat equation and it can be reduced to the heat equation by a clever change of variables (see Wilmott 1993). This is a well-documented equation and classical difference schemes approximate it well. The challenge, however, is to devise finite difference schemes that will work under all conditions, regardless of the parameters in equation (3.12).

Thus, we conclude by noting that the above discussion can be applied to two-factor models of the form

(3.19) $$-\frac{\partial u}{\partial t} + \sigma_1 \cdot \frac{\partial^2 u}{\partial x^2} + \sigma_2 \cdot \frac{\partial^2 u}{\partial y^2} + \mu_1 \cdot \frac{\partial u}{\partial x} + \mu_2 \cdot \frac{\partial u}{\partial y} + bu = f$$

This equation occurs when modeling two-factor Gaussian term structures (see Levin 2000). In this case $\sigma_1$ is the short-rate volatility constant, $\sigma_2$ is the long yield volatility and $\mu_1, \mu_2$ are given in terms of other known parameters. The reduced form is given by

(3.20) $$-\frac{\partial u}{\partial t} + \mu_1 \cdot \frac{\partial u}{\partial x} + \mu_2 \cdot \frac{\partial u}{\partial y} + bu = f$$

with the resulting loss of boundary conditions. Equation (3.20) is called hyperbolic (in the sense of Friedrichs 1958). Classical difference schemes will be no better at approximating equations (3.19) and (3.20) than in the one-dimensional case; again, robust methods are needed.

**Risk Management Motivation**

We shall now deal with numerical methods for what might be called the "time-dependent" Black & Scholes equation. This is the Black & Scholes equation after removal of the time dependencies, which, in effect translates to a two-point boundary value problem. There are three reasons for adopting this approach:

1. It is easier to solve than the full blown Black & Scholes and we already know why classical difference schemes do not always work when the coefficient of the first derivative is large. In particular, we can predict when bounded and unbounded oscillations occur in the approximate solution.
2. We can introduce a class of "fitted" difference schemes that are unconditionally stable and converge to the exact solution of the corresponding two-point boundary problem.
3. We can then transform the "fitted" schemes to solve the full Black & Scholes equation by *discretization* in the direction of the underlying $S$ (known as the semi-discrete scheme) and then in the time direction (employing time-stepping methods such as Crank-Nicholson and Runge-Kutta).

The resulting scheme is unconditionally stable and convergent and there are no oscillations to be found (for the full proof see Duffy 1980). This means that the method works without restriction to the size of the volatility, the value of the underlying or interest rate.

## Some Classical Finite Difference Schemes

Our aim is to approximate (3.12) - (3.14) by finite difference schemes. To this end we divide the interval $[A, B]$ into sub-intervals

$$A = x_0 < x_1 < \cdots < x_j = B$$

and we assume for convenience that the mesh-points $\{x_j\}_{j=0}^{J}$ are equidistant. That is

$$x_j = x_{j-1} + h, \quad j = 1, \ldots, J \left( h = \frac{B-A}{J} \right)$$

We then divide interval $[0, T]$ into $N$ equal sub-intervals

$$0 = t_0 < t_1 < \cdots < t_N = T$$

where

$$t_1 = t_{n-1} + k, \quad n = 1, \ldots, N \left( k = \frac{T}{N} \right)$$

The essence of the finite difference approach lies in replacing the derivatives in (3.12) by divided differences at the mesh points $(x_j, t_n)$. We define the difference operators in the $x$-direction as follows:

$$D_+ u_j = (u_{j+1} - u_j)/h, \quad D_- u_j = (u_j - u_{j-1})/h$$
$$D_0 u_j = (u_{j+1} - u_{j-1})/2 \cdot h, \quad D_+ D_- u_j = (u_{j+1} - 2 \cdot u_j + u_{j-1})/h^2$$

One can show by means of the Taylor Expansion that $D_+$ and $D_-$ are first order approximations to $\frac{\partial}{\partial x}$ while $D_0$ is a second order approximation to $\frac{\partial}{\partial x}$. Finally, $D_+ D_-$ is a second order approximation to $\frac{\partial^2}{\partial x^2}$.

We now consider the following two-point boundary value problem:

Find a function $u$ such that

(4.1)
$$\sigma \cdot \frac{d^2u}{dx^2} + 2 \cdot \frac{du}{dx} = 0 \quad \text{in } (0,1)$$
$$u(0) = 1, u(1) = 0$$

We assume, for the moment, that $\sigma$ is a positive constant. We now replace the derivatives in (4.1) by their corresponding finite differences in order to produce the following finite difference method:

Find a mesh-function $\{U_j\}_{j=1}^{J-1}$ such that

(4.2)
$$\sigma \cdot D_+D_-U_j + 2 \cdot D_0U_j = 0, \quad j = 1, \ldots, J-1$$
$$U_0 = 1, U_J = 0$$

Scheme (4.2) is sometimes referred to as the divided difference scheme and it is a standard way to approximate convection-diffusion equations. In particular, many risk analysts use it to approximate the Black & Scholes equation. We now show that (4.2) does not always produce correct results.

It is obvious that the exact solution to (4.1) is given by

(4.3)
$$u(x) = \frac{e^{-2 \cdot x/\sigma} - e^{-2/\sigma}}{1 - e^{-2/\sigma}}$$

and that the exact solution to (4.2) is

(4.4)
$$U_j = \frac{\lambda^j - \lambda^J}{1 - \lambda^J} \quad \text{where} \quad \lambda = \frac{1 - h/\sigma}{1 + h/\sigma}$$

(see Farrell 2000 for details).

Let us now assume that $\sigma < h$ in equation (4.4). This means that $\lambda < 0$ and thus $\lambda^j$ is positive or negative depending on whether $j$ is even or odd!

Furthermore,

$$\lim_{\sigma \to 0} U_j = \left((-1)^j + 1\right)/2$$

and as a result, $U_j$ oscillates in a bounded fashion for all $\sigma$ satisfying $\sigma < h$. It is thus, a simple matter to conclude that centered difference schemes are unsuitable for the numerical solution of problem (4.1) when $\sigma < h$. It can also be shown that this difference scheme produces a solution that goes to infinity as $\sigma \to 0$ (see Farrell page 18).

An alternative to centered divided differences could be used to approximate (4.1) by upward finite difference schemes:

(4.5)
$$\sigma D_+ D_- U_j + 2 \cdot D_+ \cdot U_j, \quad j = 1, \ldots, J-1$$
$$U_0 = 1, U_J = 0$$

The solution to (4.5) is given by

$$U_j = \frac{\lambda - \lambda^J}{1 - \lambda^J}, \lambda = \frac{1}{1 + \frac{2 \cdot h}{\sigma}}$$

We note that

$$U_1 - u(x_1) = \frac{\lambda - \lambda^J}{1 - \lambda} - \frac{r - r^J}{1 - r^J}, r = e^{-2 \cdot h / \sigma} S$$

where $u = u(x)$ is the solution defined by (4.5). If we now set $\frac{\sigma}{h} = 1$ as $N \to \infty$ we get

$$U_1 - u(x_1) = \frac{1}{3} - e^{-2} = 0.197998$$

This means that the point-wise error is ~ 20% regardless of the size of $h$. This is unacceptable. Thus, we may conclude that center divided and one-sided difference schemes will result in either spurious oscillations and/or an inaccurate solution.

## A New Class of Robust Difference Schemes

Let us now introduce the class of exponentially fitted schemes for general two-point boundary value problems and then apply these schemes to the Black & Scholes equation.

Exponentially fitted schemes are stable, possess adequate convergence properties and do not produce spurious oscillations.

A very general two point boundary value problem is defined as

(4.6)
$$\sigma \frac{d^2 u}{dx^2} + \mu \frac{du}{dx} = 0 \quad \text{in} \quad (A, B)$$
$$u(A) = \beta_0, U_J = \beta_1$$

Here we assume that $\sigma$ and $\mu$ are positive constants. Now, we can approximate (4.6) by

(4.7)
$$\sigma \rho D_+ D_- U_j + \mu D_0 U_j = 0, \quad j = 1, \ldots, J-1$$
$$U_0 = \beta_0, U_J = \beta_1$$

where $\rho$ is a fitting factor -- this factor is precisely 1 in the case of the centered difference scheme in (4.2). We now choose $\rho$ such that the solutions to (4.6) and (4.7) are identical at mesh points. Some elementary arithmetic shows that

$$\rho = \frac{\mu \cdot h}{2 \cdot \sigma} \coth \frac{\mu \cdot h}{2 \cdot \sigma}$$

where coth is the hyperbolic co-tangent function defined as

$$\coth x = \frac{e^x + e^{-x}}{e^x - e^{-x}} = \frac{e^{2x} + 1}{e^{2x} - 1}$$

The fitting factor will be used when developing fitted difference schemes for even more general two point boundary value problems. In particular, let us address the following problem

(4.8)
$$\sigma(x)\frac{d^2 u}{dx^2} + \mu(x)\frac{du}{dx} + b(x) \cdot u = f(x)$$
$$u(A) = \beta_0, u(B) = \beta_1$$

where $\sigma, \mu$ and $b$ are given continuous functions, and

$$0 \leq \sigma(x), \mu(x) \geq \alpha > 0, b(x) \leq 0 \text{ for } x \in (A, B)$$

The fitted difference scheme that approximates (4.8) is given by

(4.9)
$$\gamma_j^h D_+ D_- U_j + \mu_j D_0 U_j + b_j U_j = f_j, \quad j = 1, \ldots, J-1$$
$$U_0 = \beta_0, U_J = \beta_1$$

where

(4.10)
$$\gamma_j^h = \frac{\mu_j h}{2} \coth \frac{\mu_j h}{2\sigma_j}$$
$$\sigma_j = \sigma(x_j), \mu_j = \mu(x_j), b_j = b(x_j).$$

We will proceed to demonstrate some of the fundamental results (see Il'in 1969, Doolan 1980, Duffy 1980).

**Theorem 4.** *(Uniform Stability)*

*The solution of scheme* (4.9) *is uniformly stable, that is*

$$|U_j| \leq |\beta_0| + |\beta_1| + \frac{1}{\alpha} \max_{k=1,\ldots,J} |f_k|, \quad j = 1,\ldots,J-1$$

*What is more, scheme* (4.9) *is monotone in the sense that the matrix representation of* (4.9)

$$AU = F$$

*where* $U = {}^t(U_1,\ldots,U_{J-1})$, $F = {}^t(f_1,\ldots,f_{J-1})$ *and*

$$\mathbf{A} = \begin{pmatrix} \ddots & & \ddots & & 0 \\ & \ddots & & a_{j,j+1} & \\ \ddots & & a_{j,j} & & \ddots \\ & a_{j,j-1} & & \ddots & \\ 0 & & \ddots & & \ddots \end{pmatrix}$$

(4.11)
$$a_{j,j-1} = \frac{\gamma_j^h}{h^2} - \frac{\mu_j}{2h} > 0 \quad \text{always}$$

$$a_{j,j} = \frac{2\gamma_j}{h^2} + b_j < 0 \quad \text{always}$$

$$a_{j,j+1} = \frac{\gamma_j}{h^2} - \frac{\mu_j}{2h} > 0 \quad \text{always}$$

*produces positive solutions from positive input.*

Sufficient conditions for a difference scheme to be monotone have been investigated by many authors over the past 30 years. We draw attention to the work of Samarski 1976 and Stoyan 1979.

The latter has experimented with several fitting factors:

(4.12)
$$\rho_0 = \sigma^{-1}\{1 + q^2/(1+|q|)\}$$
$$\rho_1 = (1+q^2)^{1/2}$$
$$\rho_2 = \sigma^{-1} \cdot (\gamma)$$

where $\gamma$ is the fitting factor in (4.10).

**Theorem 5.** *(Uniform Convergence)*

*Let $u$ and $U$ be solutions of* (4.8) *and* (4.9) *respectively. Then*

$$\left| u(x_j) - U_j \right| \leq Mh$$

*Where $M$ is a positive constant that is independent of $h$ and $\sigma$.*

The conclusion is that fitted scheme (4.9) is stable, convergent and produces no oscillations for all parameter regimes. In particular the scheme degrades gracefully to a well-known stable scheme when $\sigma \to 0$.

# Conclusion

In a nutshell: we can calculate the underlying asset value and the present value of the payoff (the death benefit, which represents the only true payoff)-see 2. We can use a robust method as described in 1 to estimate volatility while side-stepping the danger of naïve assumption and simultaneously incorporating the fat-tailed traits of the underlying distribution. The risk-free rate can be estimated from the market. The process of simulation given in 3 and 4 allow us to generate paths to payoff and also provide a variance for the value that is one of perhaps many answers to our question: how much is the possible return on the mortality bet worth today?